\documentclass{aip-cp}

\usepackage[numbers]{natbib}
\usepackage{rotating}
\usepackage{graphicx}

\begin{document}

% Title portion
\title{Status of the Schwarzchild-Couder Medium-Sized Telescope for the
  Cherenkov Telescope Array}

\author[aff1]{W. Benbow\corref{cor1}}
\affil[aff1]{Harvard-Smithsonian Center for Astrophysics, 60 Garden
  St, Cambridge, MA, 02180, USA}
\author[aff2]{A. N. Otte}
\affil[aff2]{School of Physics \& Center for Relativistic Astrophysics, Georgia Institute of Technology, 837
State Street NW, Atlanta, GA 30332-0430, USA}
\author[aff3]{for the pSCT Consortium}
\affil[aff3]{http://cta-psct.physics.ucla.edu/index.html}
\author[aff4]{the CTA Consortium}
\affil[aff4]{http://cta-observatory.org/}
\corresp[cor1]{Corresponding author: wbenbow@cfa.harvard.edu}

\maketitle

\begin{abstract}
The Cherenkov Telescope Array (CTA) is planned to be the
next-generation very-high-energy (VHE; E$>$100 GeV) gamma-ray
observatory.   It is anticipated that CTA will improve upon the
sensitivity of the current generation of VHE experiments, such as
VERITAS, HESS and MAGIC, by an order of magnitude.  CTA is planned to 
consist of two graded arrays of Cherenkov telescopes with three primary-mirror sizes.
A proof-of-concept telescope, based on the dual-mirror
Schwarzchild-Couder design, is being constructed on the VERITAS site
at the F.L. Whipple Observatory in southern Arizona, USA, and is a candidate design for the
medium-sized telescopes.  The telescope's construction will be completed in early
2017, and the status of this project is presented here.
\end{abstract}

% Head 1
\section{INTRODUCTION}

The next-generation VHE observatory, CTA \cite{CTA_design_concept}, is expected to provide 
coverage between $\sim$20 GeV and $\sim$300 TeV, and improve upon the
sensitivity of the current generation of VHE experiments by an
order of magnitude.  CTA should detect hundreds of new VHE sources,
and enable studies of the VHE sky with unprecedented angular,
spectral and temporal resolution.  In addition to providing unique new
capabilities for VHE astronomy, to ensure the maximum scientific
output of the facility, the CTA consortium will operate the facility
as an open observatory, providing access to data in
the VHE band to members of the wider astronomical community for
the first time.

The current baseline design of CTA consists of two arrays of Cherenkov
telescopes located in the Southern and Northern Hemispheres.  The
arrays will include telescopes of small (D$\sim$4 m; southern site
only),  medium (D$\sim$12 m) and large sizes  (D$\sim$23 m), which
focus on providing coverage for different parts of the VHE band.  
Historically Cherenkov telescopes were constructed with segmented,
single-mirror designs of either Davies-Cotton (DC; effectively spherical)
or parabolic shape.  However, for CTA several telescopes with dual-mirror designs are
being prototyped for use as the small- and medium-sized telescopes, 
in addition to those with classical designs.  Although these dual-mirror
systems are more complex and have significantly tighter alignment 
requirements, they offer a number of advantages.  The
most-significant of these advantages is their improved optical
performance across their field of view compared to DC telescopes.  The Schwarzchild-Couder design
provides a much reduced plate scale, and a uniform optical
point-spread function  (i.e. unaffected by
spherical or comatic aberrations) over a wide field of
view (FoV).  The smaller plate scale enables the use of compact cameras for
imaging this wide FoV, which in turn enables the use of integrated, higher-efficiency photo-sensors,
and ultimately results in a much higher resolution camera.  Because
the higher resolution camera allows a more precise reconstruction and
characterization Cherenkov images, an array of dual-mirror telescopes
should outperform an array of comparably-sized single-mirror
instruments.

\begin{figure}[]
   \centerline{ {\includegraphics[width=3.5in]{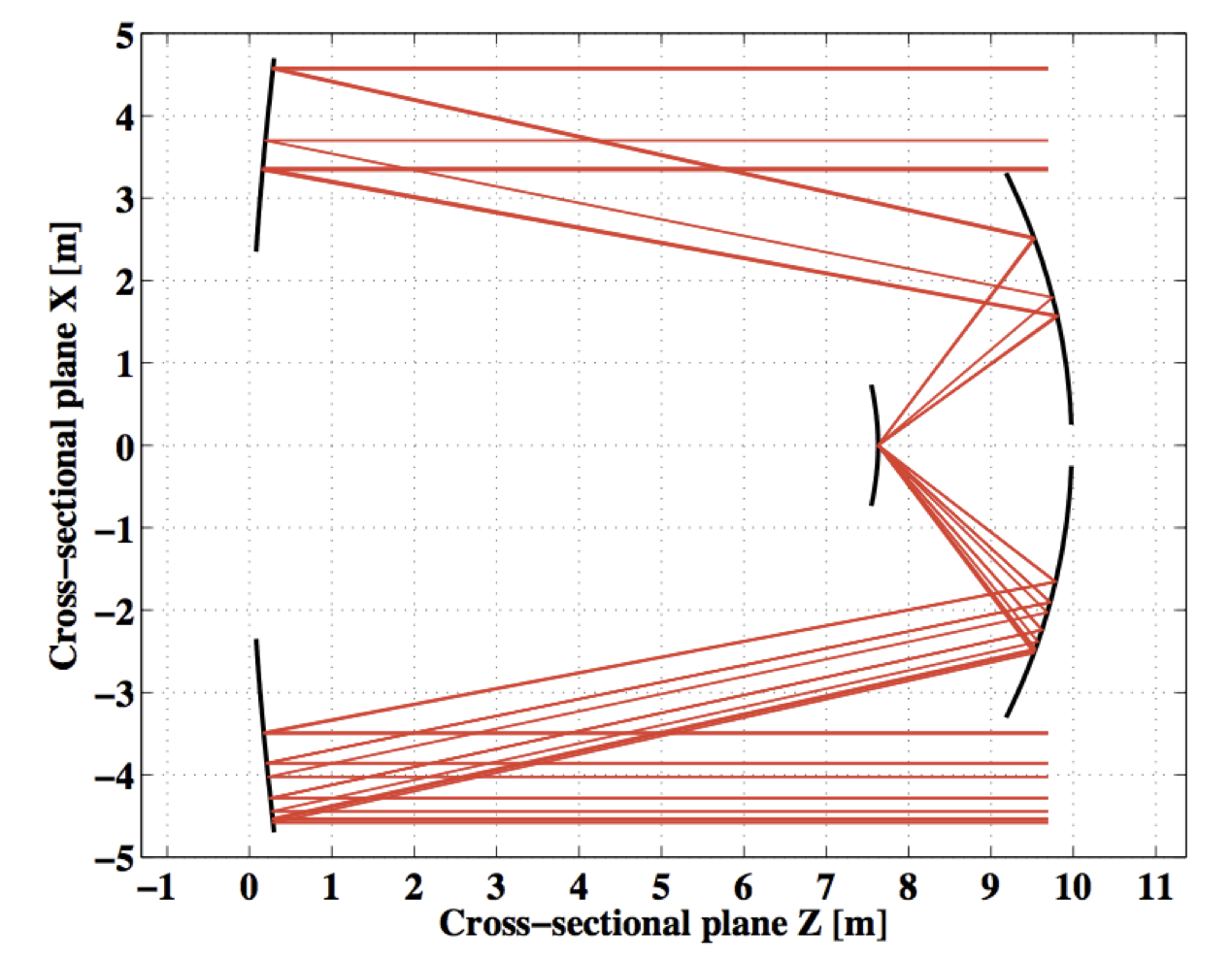} }
              \hfil
              {\includegraphics[width=2.5in]{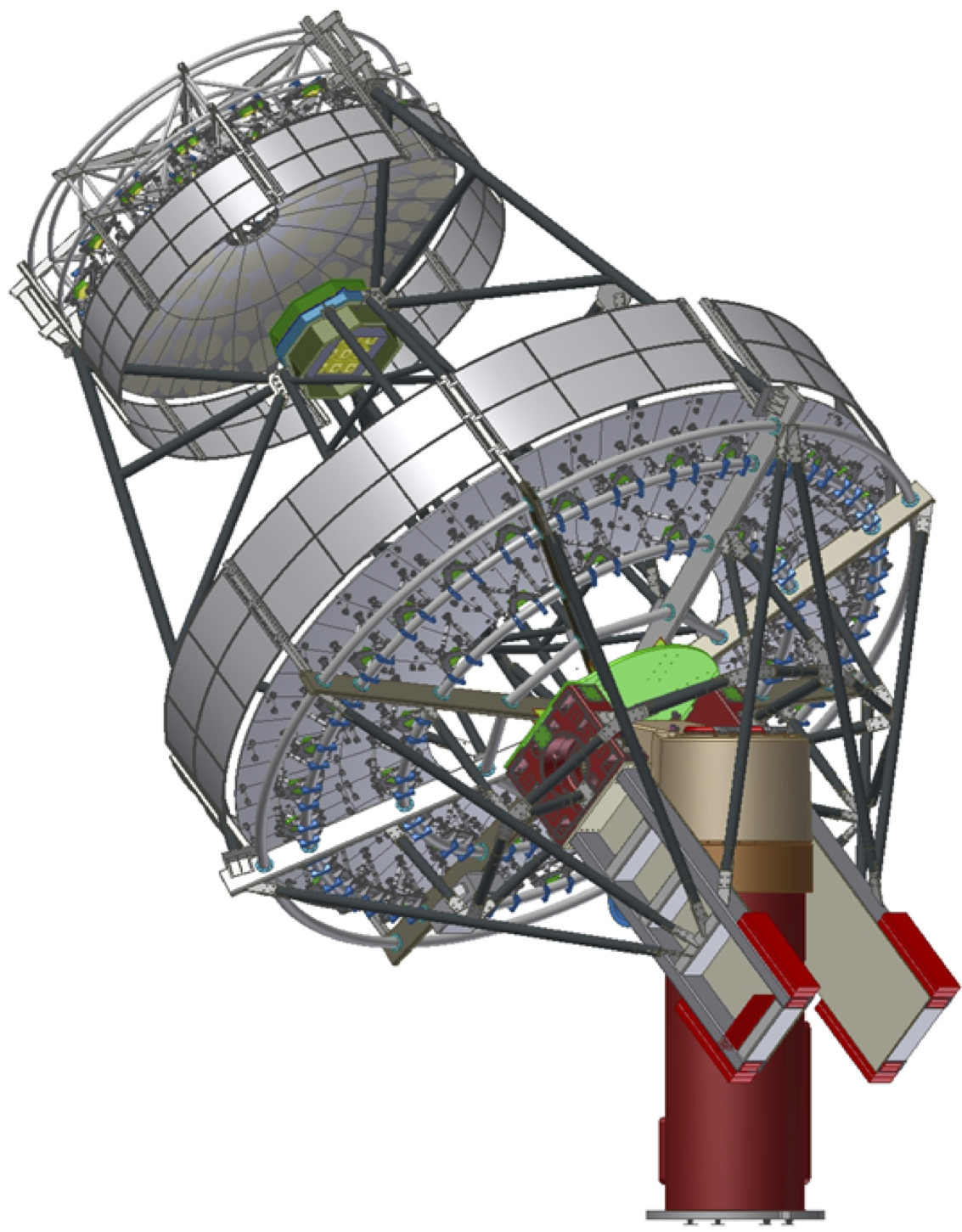} }
             }
  \caption{Left:  Schematic for the optical design of the Schwarzchild-Couder
  telescope.  Right:  Conceptual drawing of the proof-of-concept SCT
  being constructed at the F.L Whipple Observatory in Arizona, USA. }
\label{optical_design}
\end{figure}

\section{The Optical Support Structure}
\begin{figure}[]
   \centerline{ {\includegraphics[width=1.5in]{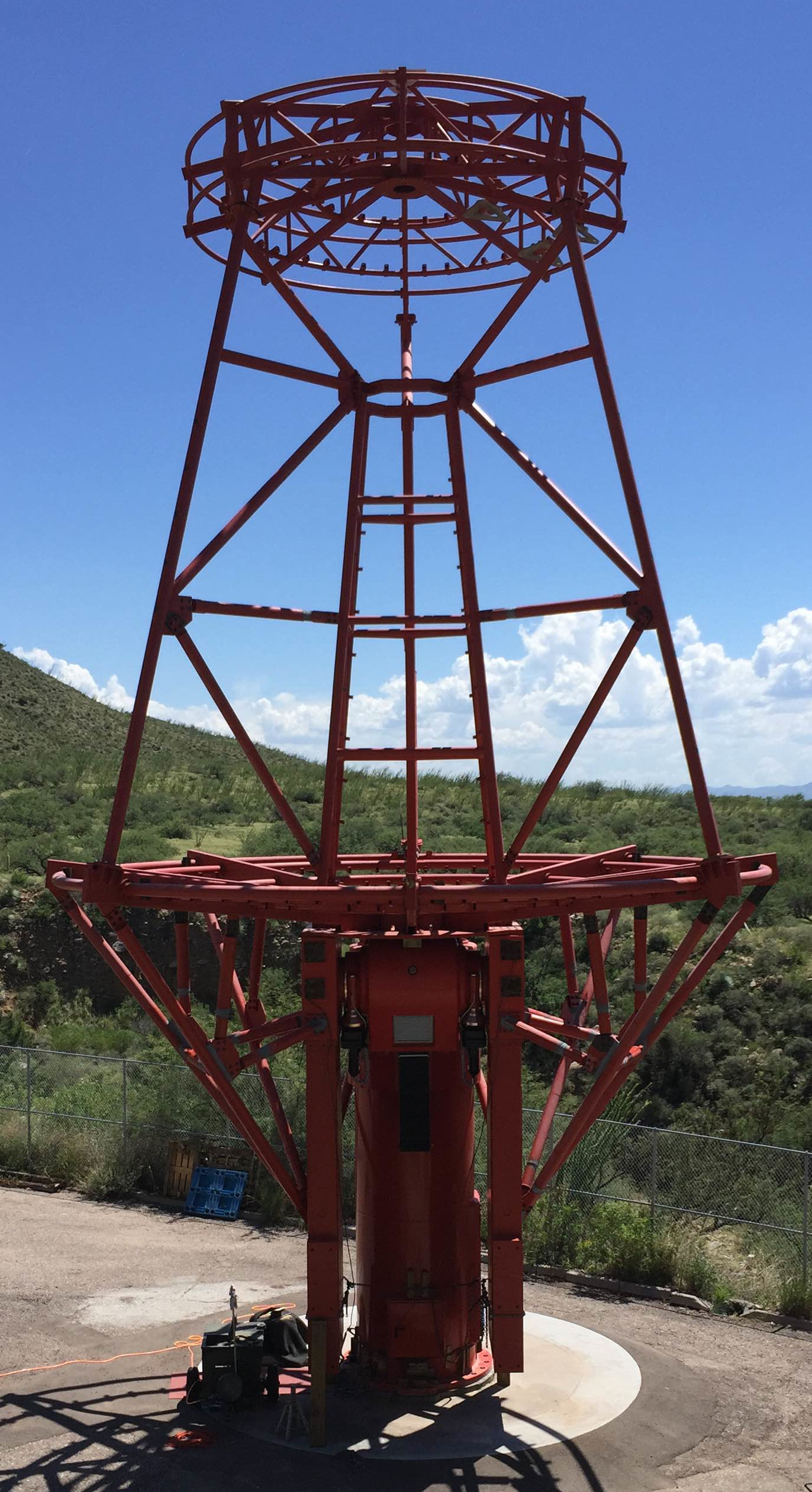} }
              \hfil
              {\includegraphics[width=1.95in]{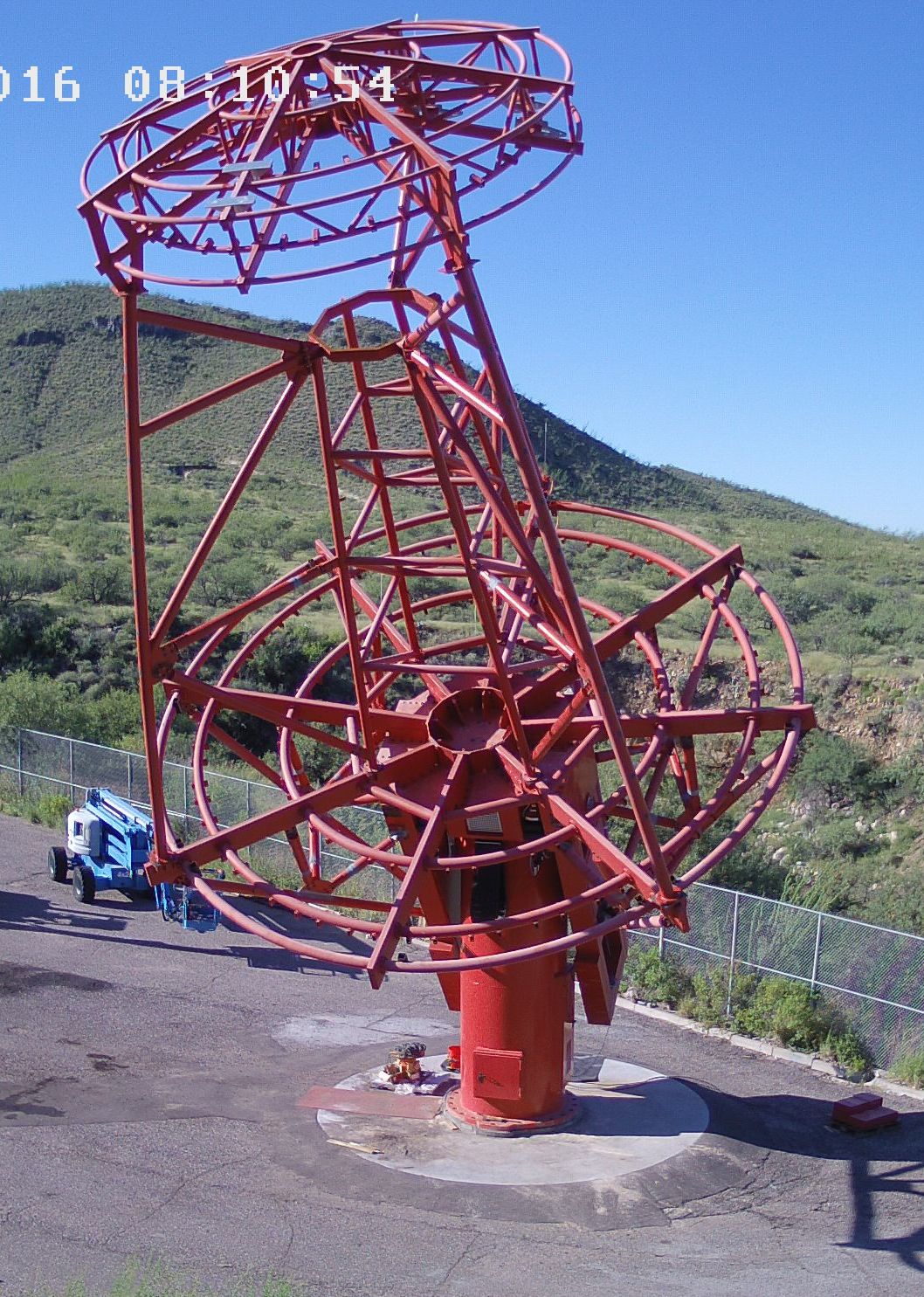} }
              \hfil
              {\includegraphics[width=2.55in]{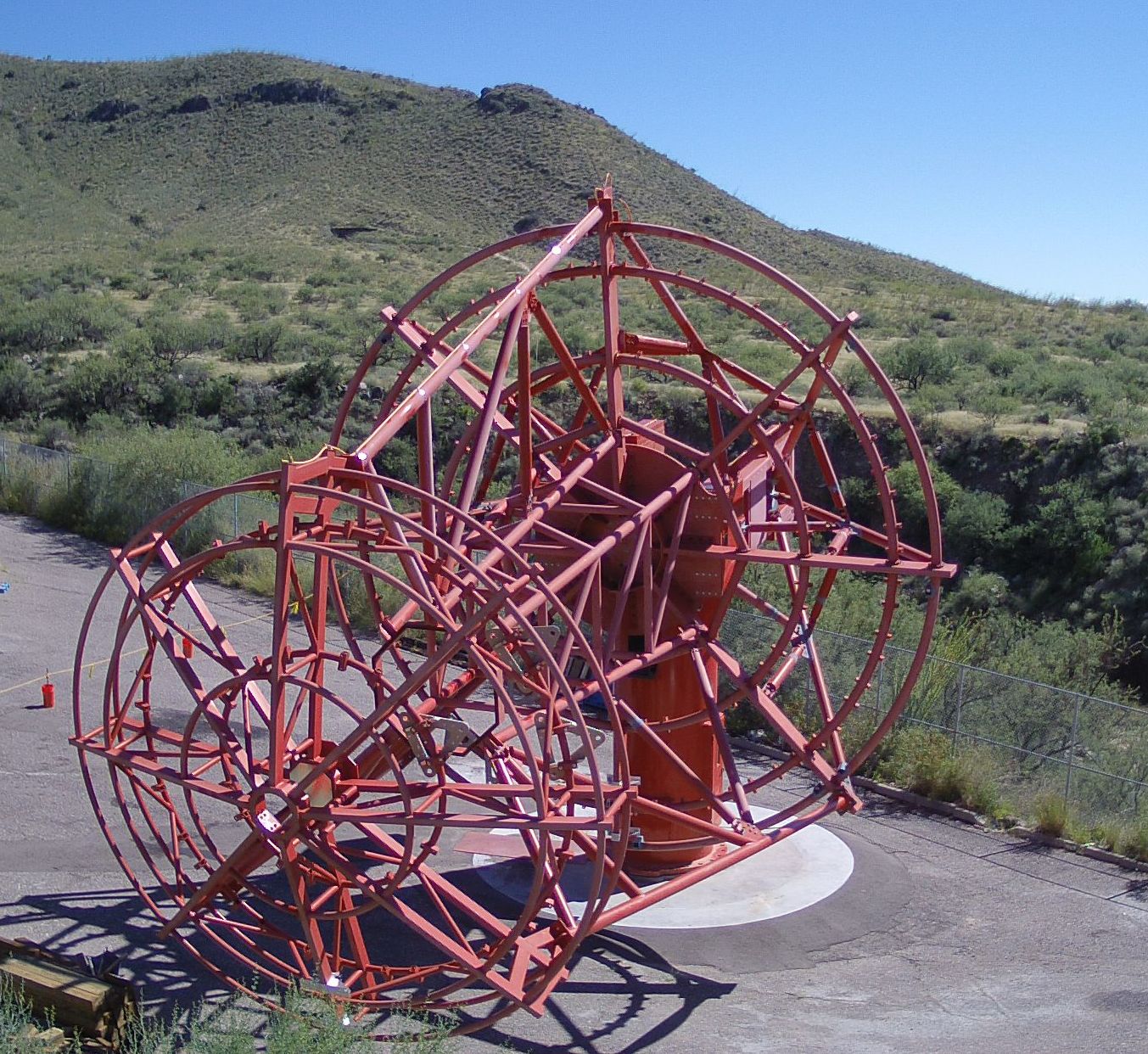} }
             }

  \caption{The completed OSS and positioner for the pSCT in September
    2016 shown at three different elevation angles.}
\label{assembled_OSS}
\end{figure}

The design of the Schwarzchild-Couder telescope (SCT) employs a novel aplanatic optical
system composed of two aspheric mirrors ~\cite{VVV_SCT_design}. Figure
~\ref{optical_design} shows the schematic of the optical design for
the SCT \cite{VVV_SCT_design}.  The
proof-of-concept, medium-sized Schwarzchild-Couder telescope being
constructed for CTA is called the pSCT, and is also shown in Figure
~\ref{optical_design} .  The pSCT will have a primary mirror of 9.7 m aperture,
a secondary mirror of 5.4 m aperture, and a focal length of 5.6 m (f/D
$\sim$ 0.58).  The telescope's steel optical support structure (OSS) will support its camera, mirrors and
auxiliary systems, and will be mounted to a main plate, along with a counterweight
structure, onto the elevation axis of a positioner composed of a head / yoke and a tower.
The pSCT's positioner is very similar to that constructed for the prototype
medium-sized telescope with Davies-Cotton design, with the major
differences being a reduced height of the tower and some technical
improvements (e.g. the addition of stow pins, higher strength bearings
in the
azimuth system, casted head / yokes, etc). The pSCT's
positioner was successfully installed in February 2016, and the successful assembly and erection
of the major components of the OSS was completed
by August 2016 (the OSS was $\sim$90\% complete at the time of the
conference).  Figure ~\ref{assembled_OSS} shows the assembled pSCT OSS
and positioner at three different elevation angles.  The assembled
system has also been driven in azimuth. The metrology of the primary
and secondary dishes was measured after their installation on the
telescope structure and is within specifications.

As can be seen in Figure~\ref{optical_design}, baffles will be placed around the primary and secondary mirror
structures, primarily to contain / eliminate reflected sunlight during the daytime for safety
reasons, but also to reduce the night sky background scatter.
These will be installed in $\sim$November 2016, after the installation
of cables and conduit on the OSS.  To eliminate reflected sunlight, it is also
required that the telescope be parked during daylight hours facing north at elevation angles that change throughout the
year.  These are -5 degrees, +20 degrees, and +45 degrees.  Stow pins,
one for each of altitude and azimuth, are used to secure the telescope
during high winds.   More details regarding the OSS and the positioner can be found in ~\cite{pSCT_structure}
and ~\cite{pSCT_positioner}, respectively. 

\section{The SCT Mirror Surface}

\begin{figure}[]
   \centerline{ {\includegraphics[width=5.0in]{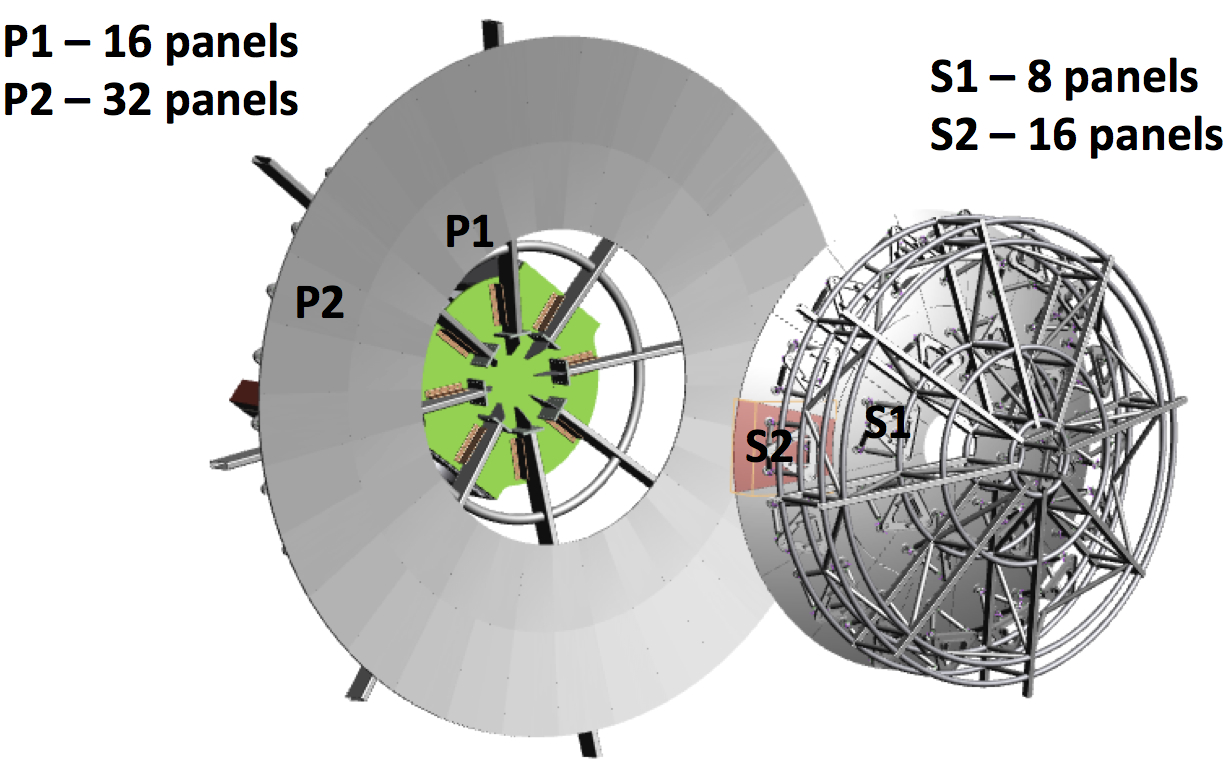} }}
  \caption{The conceptual mirror layout for the SCT.}
\label{SCT_mirror_scheme}
\end{figure}

As can be seen in Figure~\ref{SCT_mirror_scheme}, the SCT mirror
surface will consist of 72 mirror panels of 4 different shapes.  The
primary mirror surface will contain an inner ring of 16 panels, and an outer
ring of 32 panels.  The secondary mirror surface will consist of an
inner ring of 8 panels, and an outer ring of 24 panels.  Each panel is
relatively large ($\sim$1 m$^2$), and the overall primary mirror area
of $\sim$50 m$^2$ area is about half that of VERITAS or HESS.  More
details on the pSCT optical system can be found in ~\cite{pSCT_mirror_info}
After the mirrors are aligned using the equipment described in
~\cite{pSCT_mirror_align}, the optical point spread function for the 
SCT is expected to be around 5', which corresponds to a physical size of 8 mm on the focal plane.

\begin{figure}[]
   \centerline{ {\includegraphics[width=2.in]{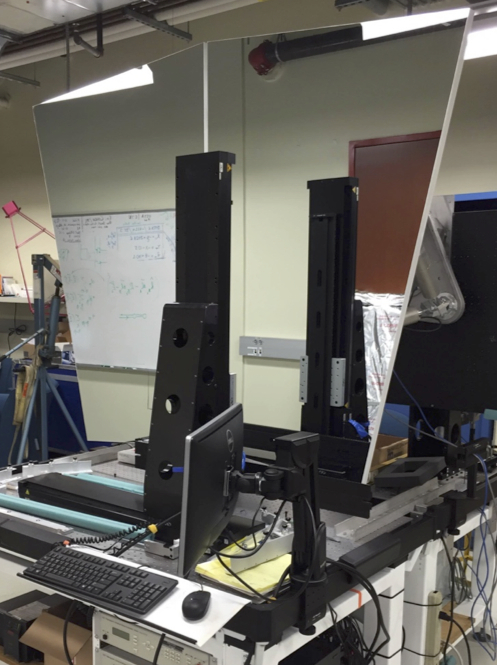} }
              \hfil
              {\includegraphics[width=3.5in]{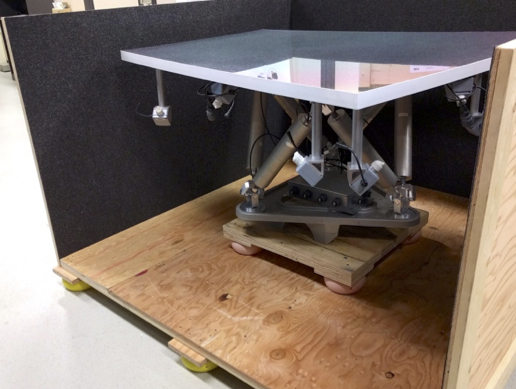} }
             }

  \caption{Left:  SCT primary mirror panels undergoing testing and
    integration at UCLA.  Right:  A fully-integrated primary-mirror
    panel ready for shipping to the pSCT site.}
\label{SCT_primary}
\end{figure}

The 48 primary mirror panels panels were produced by Media Lario Technology (MLT) in Italy using
cold-glass-slumping technology.  Here, two thin (1.7 mm) sheets of
glass are placed on each side of a 30 mm aluminum honeycomb core, and
slumped over a mandrel.  All glass panels were given a multi-layer reflective
coating by BTE in Germany and $\sim$2/3 have been delivered to the pSCT
team.  The remaining panels will be delivered in Fall 2016.
The metrology and reflectivity of the received panels were tested,
and on average are within specifications. Figure~\ref{SCT_primary}
shows two coated primary mirror panels in the UCLA testing facility. 

\begin{figure}[t]
   \centerline{ {\includegraphics[width=5.5in]{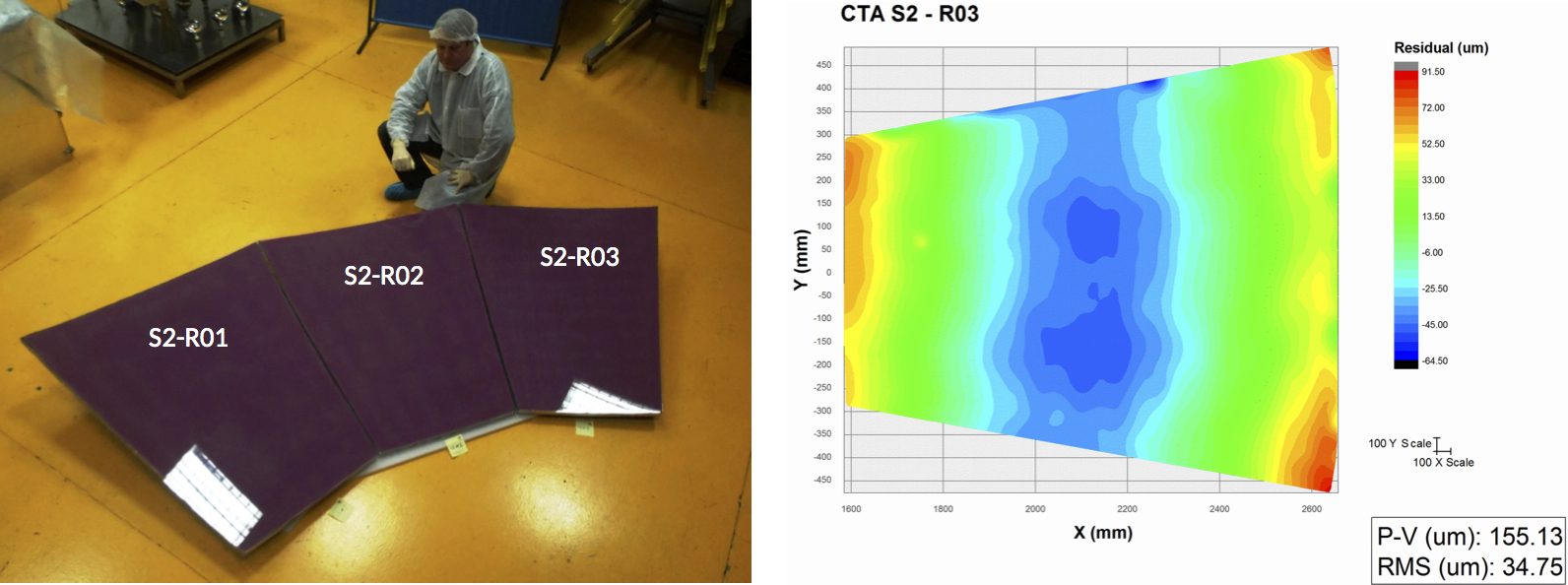} }}
  \caption{Left:  Three SCT secondary mirror panels.  Right:  Results
    of metrology testing for one of these mirror panels.}
\label{SCT_secondary}
\end{figure}

In early 2016, the fabrication of the 24 panels for the demagnifying secondary mirror surface of the pSCT
represented the main technological hurdle for the pSCT project.  A two step
process has since been successfully developed to produce these panels.  
This process includes hot slumping of flat, 12 mm float glass sheets
over a mandrel, followed by fine figuring using cold slumping
techniques similar to those for the production of the primary mirror panels.
Three panels have successfully been fabricated using this technique,
and their metrology is within specifications.
Figure~\ref{SCT_secondary} shows these panels, along with the result
of one set of metrology measurements.  Similar to the primary surface,
these panels will be coated by BTE.  Although mandrels have been
fabricated for both the inner and outer ring of secondary mirror
panels, the pSCT team is planning to produce only the outer ring of 16 panels for the initial
proof-of-concept testing.  Delivery of these panels is expected in
late 2016.  The inner ring will only be produced if sufficient resources
are identified in the future.

The primary and secondary mirror panels will all be mounted onto
independent mirror panel modules which hold and align each panel.
Each module (shown in Figure~\ref{SCT_secondary}) consists of a Stewart platform, controller board, multiple
edge sensors and a mounting triangle.    More details on these modules
and the mirror alignment systems can be found in
~\cite{pSCT_mirror_align}.  The primary mirror panels are currently
being integrated onto the mirror modules, packed into custom-made boxes, 
and will be delivered in four, sequenced shipments to the pSCT /
VERITAS site in Fall 2016.  Each module will be installed on the OSS using a crane and specialized
lifting fixture. A similar effort will commence in early 2017 for the secondary mirror panels.

\section{The Camera}

\begin{figure}[]
   \centerline{ {\includegraphics[width=3.0in]{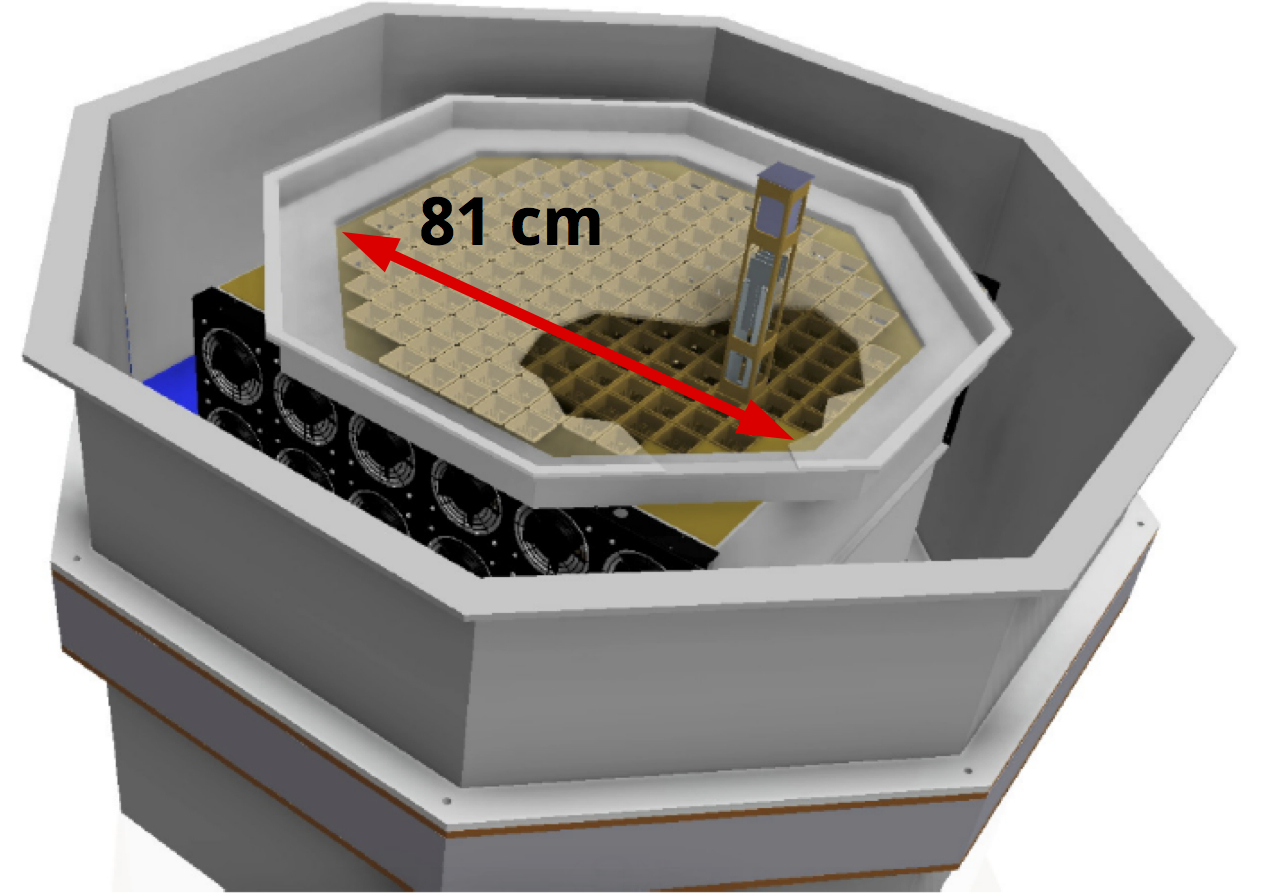} }
              \hfil
              {\includegraphics[width=3.0in]{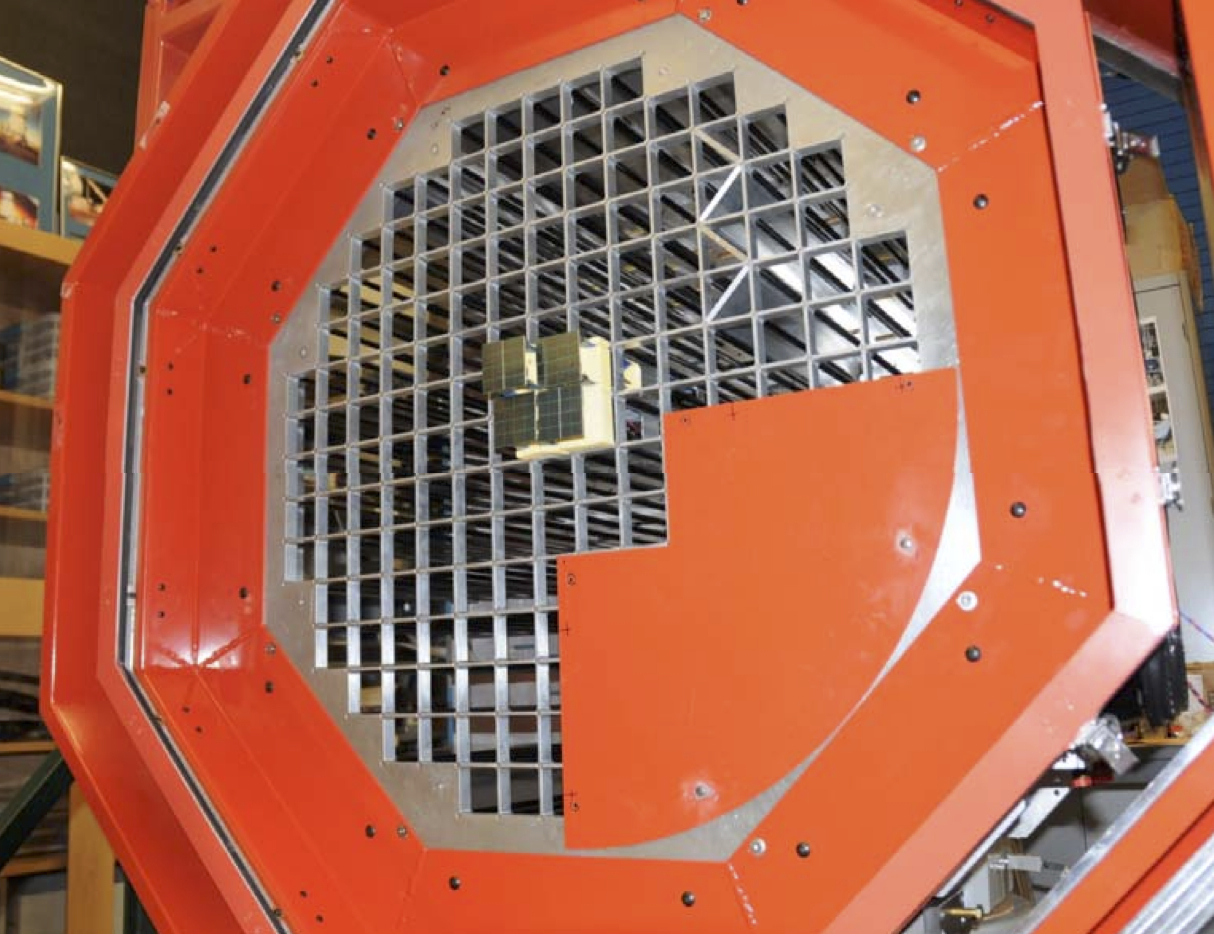} }
             }
  \caption{Left:  Conceptual design of the SCT camera.  Right:  The
    nearly complete pSCT camera.}
\label{SCT_camera_design}
\end{figure}

\begin{figure}[]
   \centerline{ {\includegraphics[width=1.05in]{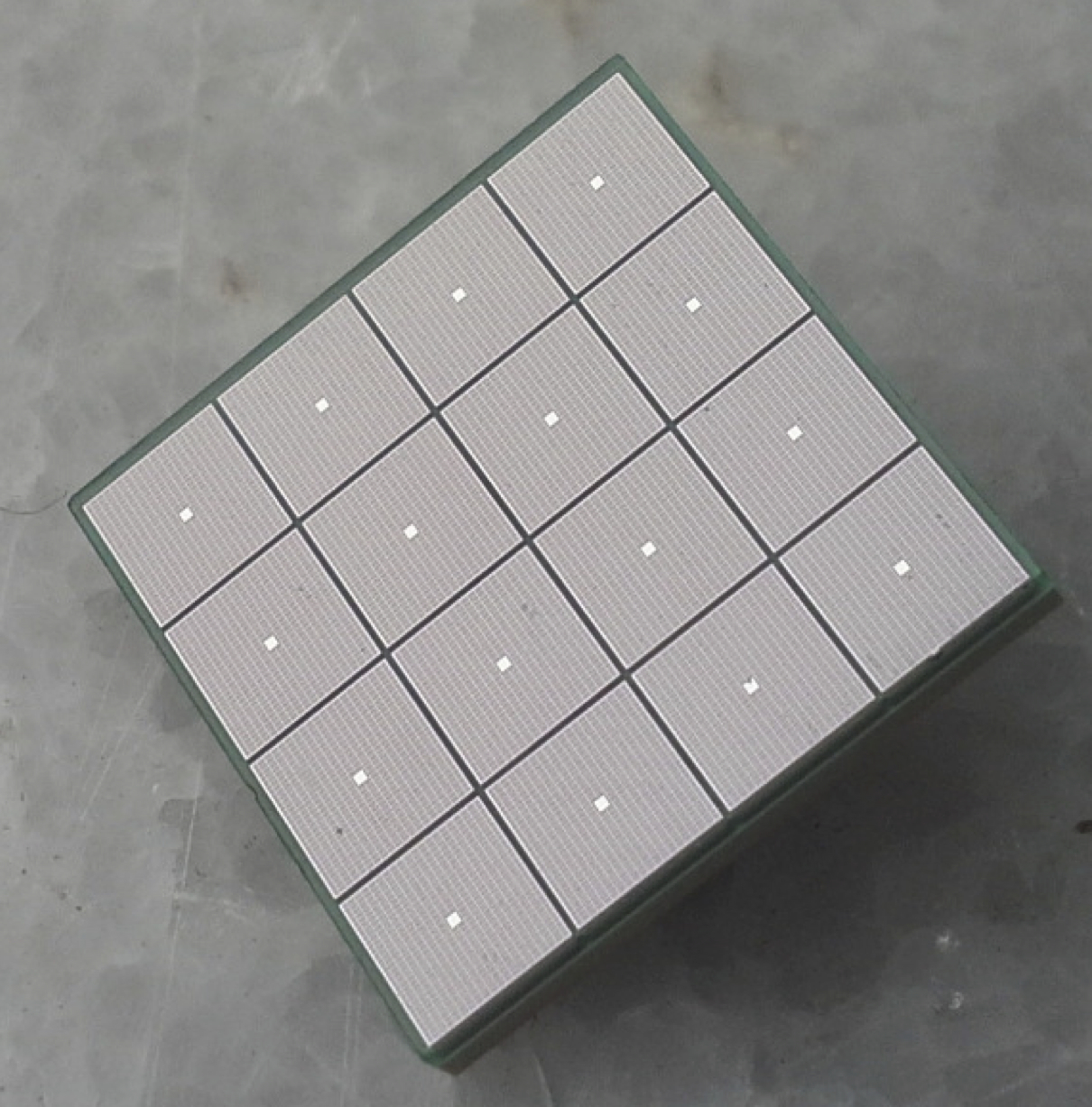} }
              \hfil
              {\includegraphics[width=4.95in]{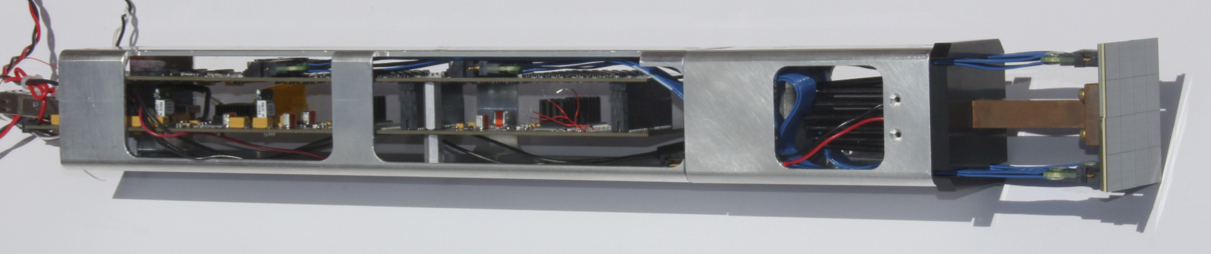} }
             }
  \caption{Left:  One tile of the Hamamatsu S12642-0404PA-50(X) used
    in the SCT prototype.  Right:  One fully assembled pSCT camera module.}
\label{SCT_camera_module}
\end{figure}

An 11,328-pixel, silicon photomultiplier (SiPM) camera was developed
for the SCT \cite{pSCT_camera}. The camera concept is shown in
Figure~\ref{SCT_camera_design}, and consists of an outer enclosure that interfaces
with the telescope structure and acts as an environmental shield for an inner
camera detector, whose surface is 81 cm width, which corresponds to
an 8$^{\circ}$ FoV.  The inner camera detector of the pSCT will use the Hamamatsu
S12642-0404PA-50(X) as the photon sensor and this device consists of
16 SiPMs (see Figure~\ref{SCT_camera_module}).  Groups of 4 SiPMs are connected in parallel to form 
each pixel with a size of 6.5 x 6.5 mm$^2$
(0.064$^{\circ}$).   The camera detector has a modular design, with each module
containing pixels on an 8 by 8 square grid of 5.4 cm width along with
their complete readout electronics.  A fully equipped SCT camera has
177 of these modules (shown in Figure~\ref{SCT_camera_module}), and the pSCT will have 16.  Although the S12642 was selected
for the pSCT, SiPM devices are rapidly improving and the SCT team is considering newer
devices for the future.  Ongoing work with FBK has shown promise, with
SiPMs already produced with similar size
to those in the pSCT and having better response in the UV/blue, better
suppression of the NSB, at least 2.5 times
lower optical cross talk, and an after-pulsing rate $<$1\%.

The complete readout electronics are contained within the SCT camera.
The front-end processing and digitization of signals, as well as the
first-level triggering are performed in the camera modules.  Nine
backplanes, on a 3 x 3 grid, handle the communication between camera
modules and servers, distribution of trigger information and power,
etc.  Of particular note is that the first-level trigger and
digitization is handled by an application-specific integrated circuit
(ASIC).  The pSCT uses the TARGET 7 chip \cite{pSCT_TARGET7}, which has
16-channels each equipped with a switched capacitor array providing
a buffer depth of $\sim$16 microseconds at a sampling rate of 1.0 GS / s.
More details regarding the pSCT data acquisition system can be found
in \cite{pSCT_DAQ}.

As of Summer 2016, the pSCT camera mechanics and electronics were
completely fabricated, and the 
remaining work consisted of TARGET calibration and module
integration which will be completed in Fall 2016.
Figure~\ref{SCT_camera_design} shows the partially populated pSCT camera.
It is anticipated that a fully-functional camera will be
delivered to the pSCT site in late 2016, and installed
on the telescope OSS using a crane and special mounting jig shortly
thereafter.

\section{Conclusions}
The dual-mirror Schwarzchild-Couder design presents a number of
challenges, but also distinct advantages for Cherenkov telescopes.  A
total of three telescopes based on this design have been prototyped
for potential use in CTA, and the pSCT is the only one of these
designs considered for the medium-sized class of telescopes.  The goal
of the pSCT project is to achieve better performance, for comparable cost, than the
slightly-larger ($\sim$12 m aperture) medium-sized Davies-Cotton
design, in particular with respect to the resolution of the recorded
Cherenkov light images.  While the SCT optics are indeed challenging,
all research and development for the project is complete and
encouraging.  The construction effort for the pSCT is ongoing 
and should be complete in early 2017, after which
the performance of the telescope can be validated.

% Acknowledgement
\section{ACKNOWLEDGMENTS}
We gratefully acknowledge support from the agencies and organizations listed under Funding
Agencies at this website: http://www.cta-observatory.org/. The development of the pSCT
has been made possible by funding provided through the U.S. National Science Foundation Major Research Instrumentation program.
% References
\nocite{*}
\bibliographystyle{aipnum-cp}%
\bibliography{pSCT_status}%

\end{document}